\documentclass{article}

\usepackage{microtype}
\usepackage{graphicx}
\usepackage{subfigure}
\usepackage{booktabs} 

\usepackage{hyperref}

\usepackage[accepted]{icml2023}

\usepackage{amsmath}
\usepackage{amssymb}
\usepackage{mathtools}
\usepackage{amsthm}
\usepackage[noend]{algpseudocode}

\usepackage[capitalize,noabbrev]{cleveref}

\theoremstyle{plain}

\theoremstyle{definition}

\theoremstyle{remark}

\usepackage[textsize=tiny]{todonotes}

\newcommand{\bx}{\mathbf{x}}
\newcommand{\by}{\mathbf{y}}
\newcommand{\bz}{\mathbf{z}}
\newcommand{\bg}{\mathbf{g}}
\newcommand{\data}{\mathcal{D}}
\newcommand{\like}{\mathcal{L}}
\newcommand{\prior}{\mathcal{P}}
\newcommand{\bH}{\mathbf{H}}
\newcommand{\bF}{\mathbf{F}}

\begin{document}
\twocolumn[
\icmltitle{Spotting Hallucinations in Inverse Problems with Data-Driven Priors}
\icmlsetsymbol{equal}{*}

\begin{icmlauthorlist}
\icmlauthor{Matt L. Sampson}{yyy}
\icmlauthor{Peter Melchior}{yyy,zzz}
\end{icmlauthorlist}

\icmlaffiliation{yyy}{Department of Astrophysical Sciences, Princeton University, Princeton, USA}
\icmlaffiliation{zzz}{Center for Statistics and Machine Learning, Princeton University, Princeton, USA}
\icmlcorrespondingauthor{Matt L. Sampson}{matt.sampson@princeton.edu}

\icmlkeywords{Machine Learning, ICML}

\vskip 0.3in
]



\printAffiliationsAndNotice{\icmlEqualContribution} 

\begin{abstract}
Hallucinations are an inescapable consequence of solving inverse problems with deep neural networks. 
The expressiveness of recent generative models is the reason why they can yield results far superior to conventional regularizers; it can also lead to realistic-looking but incorrect features, potentially undermining the trust in important aspects of the reconstruction.
We present a practical and computationally efficient method to determine, which regions in the solutions of inverse problems with data-driven priors are prone to hallucinations. 
By computing the diagonal elements of the Fisher information matrix of the likelihood and the data-driven prior separately, we can flag regions where the information is prior-dominated.
Our diagnostic can directly be compared to the reconstructed solutions and enables users to decide if measurements in such regions are robust for their application. Our method scales linearly with the number of parameters and is thus applicable in high-dimensional settings, allowing it to be rolled out broadly for the large-volume data products of future wide-field surveys.
\end{abstract}

\section{Introduction}
\label{intro}

Inverse problems commonly arise in the sciences whenever causal factors need to be inferred from observations. In imaging applications one usually seeks to infer a high-quality image from corrupted data. Examples include denoising, deconvolution, inpainting, super-resolution, and, in astronomy, deblending problems. 

Central to inverse problems is the (assumed) knowledge of the forward operator $f:\mathcal{X}\rightarrow\mathcal{Y}$, which maps the unknown parameters $\bx$ to idealized observations $\by$, which are then degraded by a lossy process to yield data $\data$ as observed. 
Inverse-problem solvers seek to find the minimum $\hat\bx$ of a loss function $L$, typically the negative logarithm of the likelihood $\like(\data\mid f, \bx)$ of the data given a model described by $f(\bx)$. 
In most cases, such a direct minimizer has undesirable properties, e.g. being noisier or less smooth than expected from a plausible solution. Regularizers are added to the loss function to promote desired properties such as smoothness or sparsity in some domain.
Increasingly, these regularizers are replaced by neural networks, trained on high-quality examples of the distribution of acceptable solutions \citep{Kamilov2022-ue, Lanusse2019-rx}.
Substantial progress has been made in the last few year in the field of generative neural networks that learn the distribution $\prior(\bx)$ of valid solutions \citep[e.g.][]{Kingma2013-ai,Goodfellow2014-yn, song2020score}.

We can now optimize the Bayesian log-posterior 
\begin{equation}
\label{eq:posterior}
 L(\bx)=\log\like(\bx) + \log\prior(\bx),   
\end{equation}
using a data-driven prior as regularizer.
As $L$ is now a combination of two terms, it is not obvious which aspects of the posterior minimizer $\hat\bx$ are actually determined by data, and which are filled in by the prior.
As generative networks have become more expressive, they have been found to be prone to hallucinations, i.e. the tendency to create features that are plausible but do not originate from $\data$ directly \citep{Gottschling2020-ec}.
In fact, \emph{hallucinations are the intended mechanism to create realistic solutions} beyond what can directly be inferred from the data \citep{liu2007face, wang2014comprehensive, yu2018generative}.

However, in many scientific studies it is imperative to know whether a feature is really present or merely hallucinated. Most evidently, in medical imaging the reconstruction of a volumetric model from CT scan data is used to identify the presence of tumors. Medical professionals would like to be sure of their diagnosis before recommending an invasive procedure \citep{Bhadra2021-se}.
In astronomy, we seek to infer higher-quality measurements from lower-quality data. Hallucinations can give us the appearance of higher quality but may inadvertently bias our results towards features often seen in the training data \citep{Schawinski2017-lv}.

With inverse problems continuing to be under-constrained, and neural network regularizers producing the so-far strongest solvers, hallucinations will stay with us. 
Here we ask the question: Can we find out where they happen?

\section{Methods}
\label{sec:methods}

\begin{figure*}
    \centering
    \includegraphics[width=\textwidth]{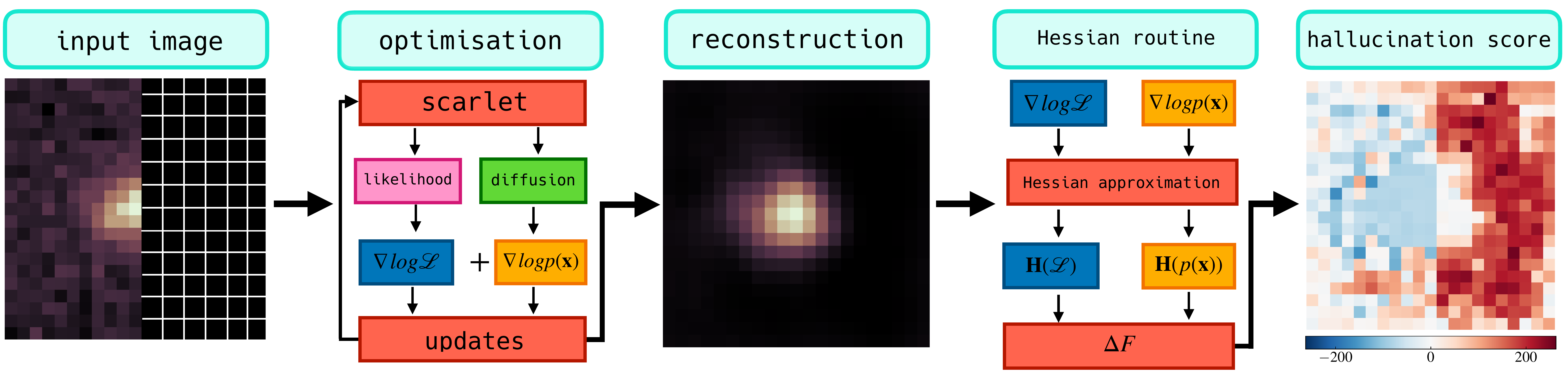}
    \caption{Overview of our methods to produce a reconstructed galaxy image and the corresponding hallucination score. The input image, of which half was masked to create an inpainting and denoising problem, is first modeled with \textsc{scarlet} by gradient descent of the likelihood $\nabla_\bx \log\mathcal{L}$ and a data-driven prior in the form of a score model $\nabla_\bx \log \mathcal{P}$. Next, we calculate the diagonalized Hessian matrices for both $\log\mathcal{L}$ and  $\log \mathcal{P}$ via \autoref{alg:cap}. We then compute the hallucination score according to \autoref{eqn:diagnostic}.}
    \label{fig:schematic}
\end{figure*}

We assume that we have access to differentiable models of the forward process $f$, the likelihood $\like$, and the prior $\prior$.
First-order minimizers then perform gradient steps in the opposite direction of  $\nabla_\bx L = \nabla_\bx \log\like + \nabla_\bx \log\prior$.
The curvature of $L$, i.e. the Hessian matrix $\bH_{ij}=\frac{\partial^2 L}{\partial x_i \partial x_j}$ is used in second-order methods to determine the step sizes.
In statistics, the negative Hessian is called the Fisher information matrix $\bF=-\bH$, so dubbed because it describes the statistical information about $\bx$ conveyed by $L$.
Equipped with $\bF$, we can recast our question: Where is $\bF$ dominated by the prior, as opposed to the likelihood? As long as the features of the solution are largely determined by $\like$, there are by definition no hallucinations.

We now seek to compute the Hessians of $\log\like$ and $\log\prior$ at the minimizer of \autoref{eq:posterior} separately, or, if more convenient, the Hessians of $L$ and one of the others, as the Fisher matrix is linear:
\begin{equation}
  \bF = \bF_{\log\like} + \bF_{\log\prior}.    
\end{equation}

Unfortunately, calculating the full Hessian matrix is inefficient, with a scaling of $\mathcal{O}(n^2)$, where $n$ is the dimension of $\mathcal{X}$. This makes computing $\bH$ intractable for most practical applications. A full marginalization to get pixel-level uncertainties would additionally require a matrix inversion.

But in imaging applications, the off-diagonal terms in the Hessian are typically small, and often confined to a few off-diagonal bands.
We therefore make the simplification of computing the Hessian diagonal $\mathbf{H}_D = \mathrm{Diag}(\mathbf{H})$.
Doing so has two advantages: There is an efficient way to compute only the diagonals of the Hessian, which we describe below; and the result has the same shape as $\bx$, which means that it has the form of an image that can be compared to the solution to indicate regions dominated by the prior.

We follow the approach of \citet{hutchinson1989stochastic,yao2021adahessian} to calculate the approximated diagonal Hessian.
Instead of computing the full Hessian, we make use of the Hessian-vector product (HVP), which can be computed efficiently with automatic differentiation.
From the chain rule, we have the following equation for the HVP, 
\begin{equation}
\label{eqn:hvp}
    \frac{\partial \bg^T \bz}{\partial \bx} = \frac{\partial \bg^T}{\partial \bx}\bz - \bg\frac{\partial \bz}{\partial \bx} = \frac{\partial \bg^T}{\partial \bx}\bz = \mathbf{H}\cdot\bz
\end{equation}
where $\mathbf{g}=\nabla_x L$, and $\bz$ is an arbitrary vector independent of $\bx$, hence $\partial \bz / \partial \bx = 0$. 
\autoref{eqn:hvp} evidently requires only $\mathcal{O}(n)$ operations, which is critical for higher-dimensional problems. To compute the diagonal approximation to the Hessian, we employ the method of \citet{hutchinson1989stochastic}:
\begin{equation}
\label{eqn:diag}
    \mathbf{H}_D = \mathbb{E}\left( \bz \odot (\mathbf{H}\cdot\bz) \right),
\end{equation}
with $\bz$ being sampled from a Rademacher distribution. 
We show an implementation of the Hessian diagonal approximation in \autoref{alg:cap}, making use gradient and Jacobian-vector product routines in \texttt{JAX}.

\begin{algorithm}[t]
\caption{Hessian diagonal approximation}
\label{alg:cap}
\begin{algorithmic}[0]
\Function{HessianDiag}{$f, \bx, \epsilon$}
\State g = \texttt{jax.grad}($f$)
\State hvp = \texttt{jax.jvp}(g, $\bx$)
\State $\bH = \texttt{jnp.zeros}(\bx.\texttt{shape})$
\State $\bH^\prime = \texttt{jnp.zeros}(\bx.\texttt{shape})$
\For{$i=0\dots$}
\State $\bz \sim \mathrm{Rademacher}(\bx.\texttt{shape})$
\State $\bH = \bH + (\bz \odot \mathrm{hvp}(\bz))$ 
\If{$\lVert \bH/(i+1) - \bH^\prime/i\rVert < \epsilon \lVert\bH/(i+1)\rVert$}
    \State \Return $\bH / (i+1)$
\EndIf
\State $\bH^\prime = \bH$
\EndFor
\EndFunction
\end{algorithmic}
\end{algorithm}

Finally, we produce our hallucination score
\begin{equation}
\label{eqn:diagnostic}
\Delta\bF = \mathrm{Diag}(\bF_{\log\prior}) - \mathrm{Diag}(\bF_{\log\like}).
\end{equation}
Equivalent forms, such as $\Delta\bF =2\,\mathrm{Diag}(\bF_{\log\prior}) - \mathrm{Diag}(\bF)$ can be chosen as well, e.g. when the Hessian of the log posterior has already been estimated during the optimization.
The positive regions of $\Delta\bF$ are prior-dominated and are therefore prone to hallucinations. The user can then decide how much trust they should place in the reconstruction of features in these regions. An example from an inpainting problem in astronomy is shown as the image on the right-hand side of \autoref{fig:schematic}.

\section{Experiments}
We demonstrate the capability of this method with a toy model of a galaxy reconstruction using the source deblending method \textsc{scarlet} \citep{melchior2018scarlet}, which computes a differentiable likelihood and performs proximal gradient descent to enforce regularization.
We replaced these regularizers by a score-based diffusion model, which directly learns $\nabla_\bx \log\prior$ from training data, and acts as an informative prior for the galaxy morphology distribution. We implement all models in \texttt{JAX} \citep{jax2018github} and \texttt{equinox} \citep{kidger2021equinox} with the diffusion model based on the implementation from \citet{song2020score}.

\subsection{Data}
The diffusion model was trained on data from the Subaru Hyper-Suprime Cam catalogue \citep{HSC_paper2018}. We extracted the existing \textsc{scarlet} models (each representing a single, isolated galaxy source) for three tracts, yielding about 600,000 examples. Doing so exploits that these models have already been deblended and deconvolved and can therefore act as examples of the true distribution of galaxy shapes. We trained the diffusion model on a single NVIDIA A100 for $1,000,000$ steps with a batch size of 256. 

\subsection{Quality of the score model}
\begin{figure}[ht]
    \includegraphics[width=0.45\textwidth]{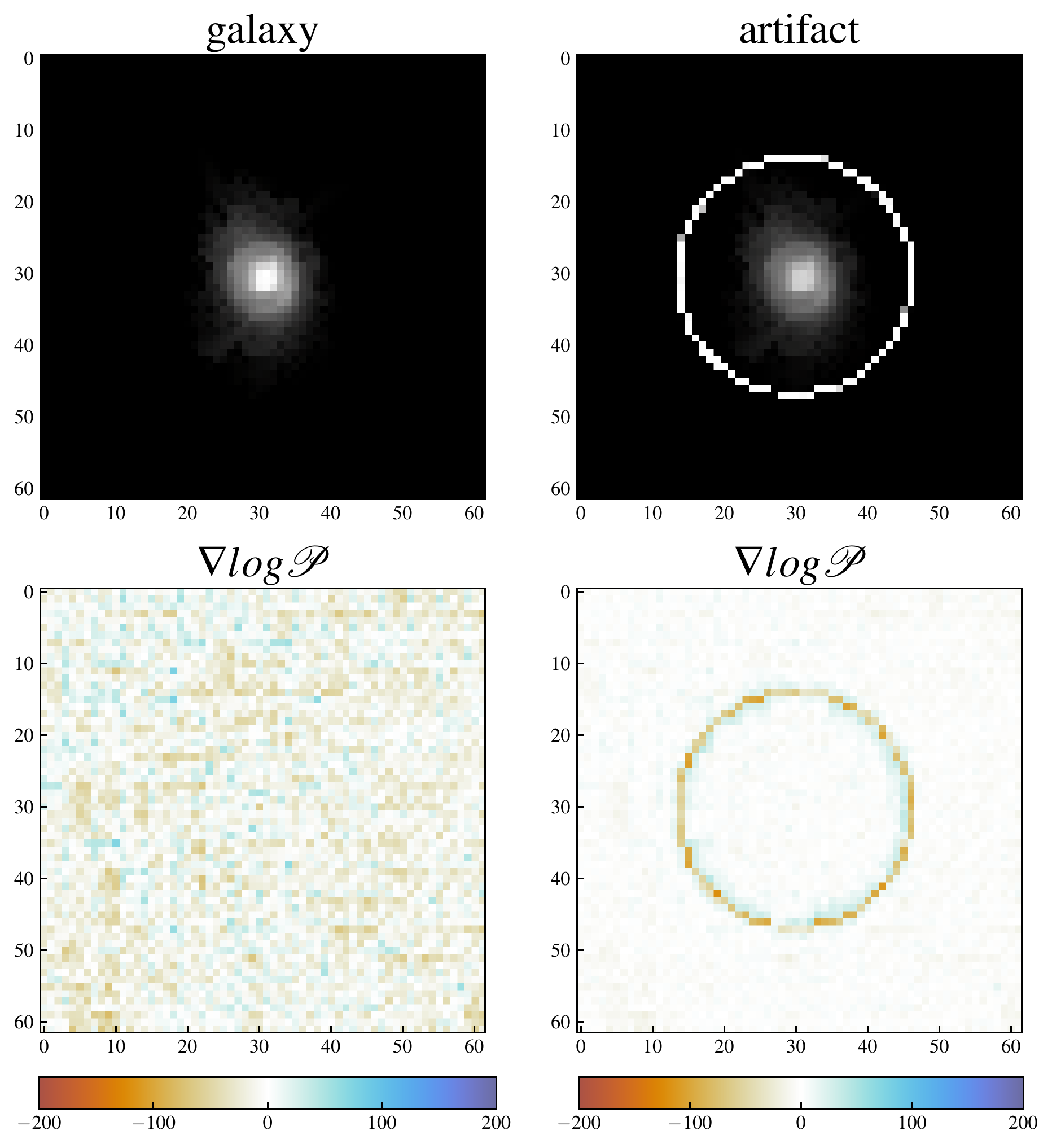}
    \caption{Example galaxy sample from the HSC catalog (\emph{top left}), with a ring artifact (\emph{top right}), and the calculated prior score for both cases (\emph{bottom row}). The artifact is strongly suppressed in the prior gradients.}
    \label{fig:artifacts}
\end{figure}

We show a test of the score model and its utility for suppressing image features that are inconsistent with galaxy shapes. \autoref{fig:artifacts} shows an input image with and without a ring-shaped artifact.
The bottom row shows the corresponding prior score.
While the original galaxy image has an overall low amplitude prior score without clear spatial structure, the artifact is strongly suppressed by the prior gradients.

While we train a time-dependent score model as in \citet{song2020score}, we evaluate the score at temperature $T=0$ during the optimization and for computing the hallucination score.
Longer runtimes of diffusion models would not allow us to scale the prior evaluations to the data volumes expected for the Vera C. Rubin Observatory \citep{Ivezic2019-bn}.
By comparing scores from the $T=0$ limit with those from full diffusion, we have confirmed that our data distribution is simple enough that the former performs sufficiently well for our purposes.
When we targeting larger or better resolved galaxies, we will need to reinvestigate this approximation.

\subsection{Hallucination score}
\begin{figure*}[t]
    \centering
    \includegraphics[width=\textwidth]{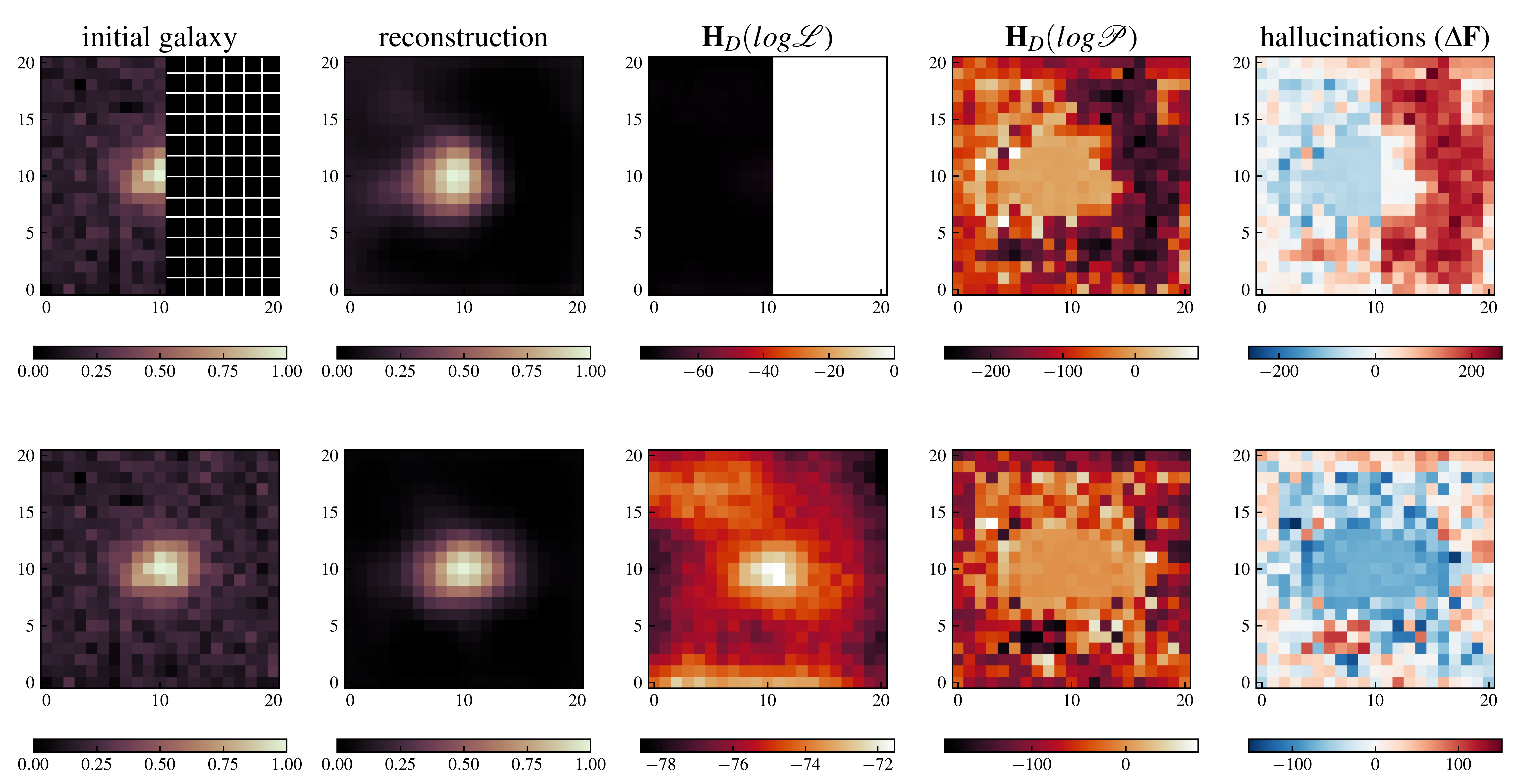}
    \caption{In the first two panels we show the initial input image for which the right half of the data has been set to 0, and the reconstructed image. Panels 3 and 4 show the Hessian diagonal for $\log \mathcal{L}$ and $\log \mathcal{P}$, respectively. We note the features of $\mathbf{H}_D (\log \mathcal{L})$ come directly from the variance weighting in the initial HSC data for the galaxy. In the rightmost, panel we show the hallucination score $\Delta \bF$ from \autoref{eqn:diagnostic}. The red shading indicates regions dominated by the prior, whereas the blue shading shows regions dominated by the likelihood.}
   \label{fig:panel5}
\end{figure*}

We assume we have a usable generative model of galaxy morphologies, which we now apply to solve an inverse problem to see where prior and likelihood dominate the reconstruction, respectively. 
We take a random sample of a galaxy observation from the HSC data and model it with \textsc{scarlet2}. For simplicity, we remove the right half of the data (array values and weights set to 0), as this will enforce the prior to generate the entirety of the features on this half of the reconstructed image. In the second trial, we perform the same reconstruction on the unaltered image. The \textbf{top row} of \autoref{fig:panel5} shows the results of trial 1, and the bottom row shows trial 2. We can see from the reconstruction that the noise in the image is removed, and that by virtue of the prior we get a reasonable estimate of the right half of the galaxy shape as well. The next panels show the Hessian diagonals $\mathbf{H}_D$ for $\log \mathcal{L}$ and $\log \mathcal{P}$, both calculated with \autoref{alg:cap}, allowing us to then compute the hallucination score from \autoref{eqn:diagnostic}. 
While the left side of the $\Delta\bF$ shows that the information comes from the likelihood, the majority of the right side is dominated by the prior, as expected due to the absence of valid data. It is noteworthy that the hallucination score is much weaker in the central region of the right side than in the outskirts. This is likely attributed to the training data consisting of primarily relatively small galaxies, leading to a high confidence in the pixel values for the outskirts of the galaxy source: they are likely very close to 0. In the inner region, the score model is much less confident, so the hallucination score is closer to 0.
We also note that for $\mathbf{H}_D(\log \mathcal{P})$ we evaluate the Jacobian of the score model, which is in itself only an approximation of the true prior gradients; any inaccuracies of the score model will be amplified when computing another derivative.

The bottom row of \autoref{fig:panel5} shows the same results run on the unaltered galaxy, where we now see that the likelihood dominates the central region with the highest signal-to-noise ratio, while the prior starts to dominate in the outskirts.

The calculations for the Hessian diagonals in the example of \autoref{fig:panel5} took $\approx$1 ms for the likelihood, which converged after a single iteration, and $\approx$260 ms for the prior, which took on average 80 iterations to converge, resulting in roughly 3 ms per single HVP evaluation.
With just-in-time compilation, the JVP of the score network is only a factor 3 slower than the HVP of the simple Gaussian likelihood of this example. All timing tests were run on an M1 Macbook Pro, utilizing 4 CPU cores.

\section{Conclusion}
We presented a practical and computationally efficient method to determine which regions in the solutions of inverse problems with data-driven priors are prone to hallucinations.
By computing the diagonal elements of the Fisher information matrix of the likelihood and the prior separately, we can flag regions where the information is prior-dominated.
Our diagnostic can directly be compared to the reconstructed solutions and enables users to make informed decisions about the trustworthiness of relevant features in the reconstruction.
Our method scales linearly with the number of parameters and is thus scalable to high-dimensional settings, allowing it to be rolled out broadly for the large-volume data products of future wide-field surveys.

The choice of \autoref{eqn:diagnostic} as a hallucination metric has advantages over simpler diagnostics such as directly calculating the standard deviation of the posterior. Doing so cannot differentiate the source of the information that determines the optimized model. Alternatively, merely checking the gradients of the likelihood and prior will become meaningless once the model is converged because they need to be either very small or cancel each other.
While our method does not make assumptions about the prior model, computing gradients of data-driven priors requires a high level of fidelity of that model.
Caution should be taken to ensure that the prior model is of sufficient accuracy for this purpose.

\section*{Software and Data}
We have used \texttt{python} \citep{van1995python} with the packages \texttt{JAX} \citep{jax2018github}, \texttt{equinox} \citep{kidger2021equinox}, \texttt{numpy} \citep{harris2020array}, and \texttt{matplotlib} \citep{Hunter:2007}. Data for this project is taken from the Subaru Hyper-Suprime Cam Survey \citep{HSC_paper2018}.

\section*{Acknowledgements}
We thank the reviewers for their helpful comments in improving this manuscript. MLS acknowledges financial support from the Princeton University First-Year Fellowship in the Natural Sciences and Engineering.

\bibliography{refs}

\begin{thebibliography}{21}
\providecommand{\natexlab}[1]{#1}
\providecommand{\url}[1]{\texttt{#1}}
\expandafter\ifx\csname urlstyle\endcsname\relax
  \providecommand{\doi}[1]{doi: #1}\else
  \providecommand{\doi}{doi: \begingroup \urlstyle{rm}\Url}\fi

\bibitem[Bhadra et~al.(2021)Bhadra, Kelkar, Brooks, and
  Anastasio]{Bhadra2021-se}
Bhadra, S., Kelkar, V.~A., Brooks, F.~J., and Anastasio, M.~A.
\newblock On hallucinations in tomographic image reconstruction.
\newblock \emph{IEEE transactions on medical imaging}, 40\penalty0
  (11):\penalty0 3249--3260, November 2021.
\newblock ISSN 0278-0062, 1558-254X.
\newblock \doi{10.1109/TMI.2021.3077857}.
\newblock URL \url{http://dx.doi.org/10.1109/TMI.2021.3077857}.

\bibitem[Bosch et~al.(2018)Bosch, Armstrong, Bickerton, Furusawa, Ikeda, Koike,
  Lupton, Mineo, Price, Takata, et~al.]{HSC_paper2018}
Bosch, J., Armstrong, R., Bickerton, S., Furusawa, H., Ikeda, H., Koike, M.,
  Lupton, R., Mineo, S., Price, P., Takata, T., et~al.
\newblock The hyper suprime-cam software pipeline.
\newblock \emph{Publications of the Astronomical Society of Japan}, 70\penalty0
  (SP1):\penalty0 S5, 2018.

\bibitem[Bradbury et~al.(2018)Bradbury, Frostig, Hawkins, Johnson, Leary,
  Maclaurin, Necula, Paszke, Vander{P}las, Wanderman-{M}ilne, and
  Zhang]{jax2018github}
Bradbury, J., Frostig, R., Hawkins, P., Johnson, M.~J., Leary, C., Maclaurin,
  D., Necula, G., Paszke, A., Vander{P}las, J., Wanderman-{M}ilne, S., and
  Zhang, Q.
\newblock {JAX}: composable transformations of {P}ython+{N}um{P}y programs,
  2018.
\newblock URL \url{http://github.com/google/jax}.

\bibitem[Goodfellow et~al.(2014)Goodfellow, Pouget-Abadie, Mirza, Xu,
  Warde-Farley, Ozair, Courville, and Bengio]{Goodfellow2014-yn}
Goodfellow, I.~J., Pouget-Abadie, J., Mirza, M., Xu, B., Warde-Farley, D.,
  Ozair, S., Courville, A., and Bengio, Y.
\newblock Generative adversarial networks.
\newblock \emph{arXiv e-prints}, June 2014.
\newblock URL \url{http://arxiv.org/abs/1406.2661}.

\bibitem[Gottschling et~al.(2020)Gottschling, Antun, Hansen, and
  Adcock]{Gottschling2020-ec}
Gottschling, N.~M., Antun, V., Hansen, A.~C., and Adcock, B.
\newblock The troublesome kernel -- on hallucinations, no free lunches and the
  accuracy-stability trade-off in inverse problems.
\newblock \emph{arXiv e-prints}, January 2020.
\newblock URL \url{http://arxiv.org/abs/2001.01258}.

\bibitem[Harris et~al.(2020)Harris, Millman, van~der Walt, Gommers, Virtanen,
  Cournapeau, Wieser, Taylor, Berg, Smith, Kern, Picus, Hoyer, van Kerkwijk,
  Brett, Haldane, del R{\'{i}}o, Wiebe, Peterson, G{\'{e}}rard-Marchant,
  Sheppard, Reddy, Weckesser, Abbasi, Gohlke, and Oliphant]{harris2020array}
Harris, C.~R., Millman, K.~J., van~der Walt, S.~J., Gommers, R., Virtanen, P.,
  Cournapeau, D., Wieser, E., Taylor, J., Berg, S., Smith, N.~J., Kern, R.,
  Picus, M., Hoyer, S., van Kerkwijk, M.~H., Brett, M., Haldane, A., del
  R{\'{i}}o, J.~F., Wiebe, M., Peterson, P., G{\'{e}}rard-Marchant, P.,
  Sheppard, K., Reddy, T., Weckesser, W., Abbasi, H., Gohlke, C., and Oliphant,
  T.~E.
\newblock Array programming with {NumPy}.
\newblock \emph{Nature}, 585\penalty0 (7825):\penalty0 357--362, September
  2020.
\newblock \doi{10.1038/s41586-020-2649-2}.
\newblock URL \url{https://doi.org/10.1038/s41586-020-2649-2}.

\bibitem[Hunter(2007)]{Hunter:2007}
Hunter, J.~D.
\newblock Matplotlib: A 2d graphics environment.
\newblock \emph{Computing in Science \& Engineering}, 9\penalty0 (3):\penalty0
  90--95, 2007.
\newblock \doi{10.1109/MCSE.2007.55}.

\bibitem[Hutchinson(1989)]{hutchinson1989stochastic}
Hutchinson, M.~F.
\newblock A stochastic estimator of the trace of the influence matrix for
  laplacian smoothing splines.
\newblock \emph{Communications in Statistics-Simulation and Computation},
  18\penalty0 (3):\penalty0 1059--1076, 1989.

\bibitem[Ivezi{\'c} et~al.(2019)Ivezi{\'c}, Kahn, Anthony~Tyson, Abel, Acosta,
  Allsman, Alonso, AlSayyad, Anderson, Andrew, Angel, Angeli, Ansari,
  Antilogus, Araujo, Armstrong, Arndt, Astier, Aubourg, Auza, Axelrod, Bard,
  Barr, Barrau, Bartlett, Bauer, Bauman, Baumont, Bechtol, Bechtol, Becker,
  Becla, Beldica, Bellavia, Bianco, Biswas, Blanc, Blazek, Blandford, Bloom,
  Bogart, Bond, Booth, Borgland, Borne, Bosch, Boutigny, Brackett, Bradshaw,
  Brandt, Brown, Bullock, Burchat, Burke, Cagnoli, Calabrese, Callahan, Callen,
  Carlin, Carlson, Chandrasekharan, Charles-Emerson, Chesley, Cheu, Chiang,
  Chiang, Chirino, Chow, Ciardi, Claver, Cohen-Tanugi, Cockrum, Coles,
  Connolly, Cook, Cooray, Covey, Cribbs, Cui, Cutri, Daly, Daniel, Daruich,
  Daubard, Daues, Dawson, Delgado, Dellapenna, de~Peyster, de~Val-Borro, Digel,
  Doherty, Dubois, Dubois-Felsmann, Durech, Economou, Eifler, Eracleous,
  Emmons, Neto, Ferguson, Figueroa, Fisher-Levine, Focke, Foss, Frank, Freemon,
  Gangler, Gawiser, Geary, Gee, Geha, Gessner, Gibson, Kirk~Gilmore, Glanzman,
  Glick, Goldina, Goldstein, Goodenow, Graham, Gressler, Gris, Guy, Guyonnet,
  Haller, Harris, Hascall, Haupt, Hernandez, Herrmann, Hileman, Hoblitt,
  Hodgson, Hogan, Howard, Huang, Huffer, Ingraham, Innes, Jacoby, Jain, Jammes,
  James~Jee, Jenness, Jernigan, Jevremovi{\'c}, Johns, Johnson, Johnson,
  Lynne~Jones, Juramy-Gilles, Juri{\'c}, Kalirai, Kallivayalil, Kalmbach,
  Kantor, Karst, Kasliwal, Kelly, Kessler, Kinnison, Kirkby, Knox, Kotov,
  Krabbendam, Simon~Krughoff, Kub{\'a}nek, Kuczewski, Kulkarni, Ku, Kurita,
  Lage, Lambert, Lange, Brian~Langton, Le~Guillou, Levine, Liang, Lim, Lintott,
  Long, Lopez, Lotz, Lupton, Lust, MacArthur, Mahabal, Mandelbaum, Markiewicz,
  Marsh, Marshall, Marshall, May, McKercher, McQueen, Meyers, Migliore, Miller,
  Mills, Miraval, Moeyens, Moolekamp, Monet, Moniez, Monkewitz, Montgomery,
  Morrison, Mueller, Muller, Arancibia, Neill, Newbry, Nief, Nomerotski,
  Nordby, O'Connor, Oliver, Olivier, Olsen, O'Mullane, Ortiz, Osier, Owen,
  Pain, Palecek, Parejko, Parsons, Pease, Matt~Peterson, Peterson, Petravick,
  Libby~Petrick, Petry, Pierfederici, Pietrowicz, Pike, Pinto, Plante, Plate,
  Plutchak, Price, Prouza, Radeka, Rajagopal, Rasmussen, Regnault, Reil, Reiss,
  Reuter, Ridgway, Riot, Ritz, Robinson, Roby, Roodman, Rosing, Roucelle,
  Rumore, Russo, Saha, Sassolas, Schalk, Schellart, Schindler, Schmidt,
  Schneider, Schneider, Schoening, Schumacher, Schwamb, Sebag, Selvy,
  Sembroski, Seppala, Serio, Serrano, Shaw, Shipsey, Sick, Silvestri, Slater,
  Allyn~Smith, Chris~Smith, Sobhani, Soldahl, Storrie-Lombardi, Stover,
  Strauss, Street, Stubbs, Sullivan, Sweeney, Swinbank, Szalay, Takacs, Tether,
  Thaler, Thayer, Thomas, Thornton, Thukral, Tice, Trilling, Turri, Van~Berg,
  Vanden~Berk, Vetter, Virieux, Vucina, Wahl, Walkowicz, Walsh, Walter, Wang,
  Wang, Warner, Wiecha, Willman, Winters, Wittman, Wolff, Michael Wood-Vasey,
  Wu, Xin, Yoachim, and Zhan]{Ivezic2019-bn}
Ivezi{\'c}, {\v Z}., Kahn, S.~M., Anthony~Tyson, J., Abel, B., Acosta, E.,
  Allsman, R., Alonso, D., AlSayyad, Y., Anderson, S.~F., Andrew, J., Angel, J.
  R.~P., Angeli, G.~Z., Ansari, R., Antilogus, P., Araujo, C., Armstrong, R.,
  Arndt, K.~T., Astier, P., Aubourg, {\'E}., Auza, N., Axelrod, T.~S., Bard,
  D.~J., Barr, J.~D., Barrau, A., Bartlett, J.~G., Bauer, A.~E., Bauman, B.~J.,
  Baumont, S., Bechtol, E., Bechtol, K., Becker, A.~C., Becla, J., Beldica, C.,
  Bellavia, S., Bianco, F.~B., Biswas, R., Blanc, G., Blazek, J., Blandford,
  R.~D., Bloom, J.~S., Bogart, J., Bond, T.~W., Booth, M.~T., Borgland, A.~W.,
  Borne, K., Bosch, J.~F., Boutigny, D., Brackett, C.~A., Bradshaw, A., Brandt,
  W.~N., Brown, M.~E., Bullock, J.~S., Burchat, P., Burke, D.~L., Cagnoli, G.,
  Calabrese, D., Callahan, S., Callen, A.~L., Carlin, J.~L., Carlson, E.~L.,
  Chandrasekharan, S., Charles-Emerson, G., Chesley, S., Cheu, E.~C., Chiang,
  H.-F., Chiang, J., Chirino, C., Chow, D., Ciardi, D.~R., Claver, C.~F.,
  Cohen-Tanugi, J., Cockrum, J.~J., Coles, R., Connolly, A.~J., Cook, K.~H.,
  Cooray, A., Covey, K.~R., Cribbs, C., Cui, W., Cutri, R., Daly, P.~N.,
  Daniel, S.~F., Daruich, F., Daubard, G., Daues, G., Dawson, W., Delgado, F.,
  Dellapenna, A., de~Peyster, R., de~Val-Borro, M., Digel, S.~W., Doherty, P.,
  Dubois, R., Dubois-Felsmann, G.~P., Durech, J., Economou, F., Eifler, T.,
  Eracleous, M., Emmons, B.~L., Neto, A.~F., Ferguson, H., Figueroa, E.,
  Fisher-Levine, M., Focke, W., Foss, M.~D., Frank, J., Freemon, M.~D.,
  Gangler, E., Gawiser, E., Geary, J.~C., Gee, P., Geha, M., Gessner, C. J.~B.,
  Gibson, R.~R., Kirk~Gilmore, D., Glanzman, T., Glick, W., Goldina, T.,
  Goldstein, D.~A., Goodenow, I., Graham, M.~L., Gressler, W.~J., Gris, P.,
  Guy, L.~P., Guyonnet, A., Haller, G., Harris, R., Hascall, P.~A., Haupt, J.,
  Hernandez, F., Herrmann, S., Hileman, E., Hoblitt, J., Hodgson, J.~A., Hogan,
  C., Howard, J.~D., Huang, D., Huffer, M.~E., Ingraham, P., Innes, W.~R.,
  Jacoby, S.~H., Jain, B., Jammes, F., James~Jee, M., Jenness, T., Jernigan,
  G., Jevremovi{\'c}, D., Johns, K., Johnson, A.~S., Johnson, M. W.~G.,
  Lynne~Jones, R., Juramy-Gilles, C., Juri{\'c}, M., Kalirai, J.~S.,
  Kallivayalil, N.~J., Kalmbach, B., Kantor, J.~P., Karst, P., Kasliwal, M.~M.,
  Kelly, H., Kessler, R., Kinnison, V., Kirkby, D., Knox, L., Kotov, I.~V.,
  Krabbendam, V.~L., Simon~Krughoff, K., Kub{\'a}nek, P., Kuczewski, J.,
  Kulkarni, S., Ku, J., Kurita, N.~R., Lage, C.~S., Lambert, R., Lange, T.,
  Brian~Langton, J., Le~Guillou, L., Levine, D., Liang, M., Lim, K.-T.,
  Lintott, C.~J., Long, K.~E., Lopez, M., Lotz, P.~J., Lupton, R.~H., Lust,
  N.~B., MacArthur, L.~A., Mahabal, A., Mandelbaum, R., Markiewicz, T.~W.,
  Marsh, D.~S., Marshall, P.~J., Marshall, S., May, M., McKercher, R., McQueen,
  M., Meyers, J., Migliore, M., Miller, M., Mills, D.~J., Miraval, C., Moeyens,
  J., Moolekamp, F.~E., Monet, D.~G., Moniez, M., Monkewitz, S., Montgomery,
  C., Morrison, C.~B., Mueller, F., Muller, G.~P., Arancibia, F.~M., Neill,
  D.~R., Newbry, S.~P., Nief, J.-Y., Nomerotski, A., Nordby, M., O'Connor, P.,
  Oliver, J., Olivier, S.~S., Olsen, K., O'Mullane, W., Ortiz, S., Osier, S.,
  Owen, R.~E., Pain, R., Palecek, P.~E., Parejko, J.~K., Parsons, J.~B., Pease,
  N.~M., Matt~Peterson, J., Peterson, J.~R., Petravick, D.~L., Libby~Petrick,
  M.~E., Petry, C.~E., Pierfederici, F., Pietrowicz, S., Pike, R., Pinto,
  P.~A., Plante, R., Plate, S., Plutchak, J.~P., Price, P.~A., Prouza, M.,
  Radeka, V., Rajagopal, J., Rasmussen, A.~P., Regnault, N., Reil, K.~A.,
  Reiss, D.~J., Reuter, M.~A., Ridgway, S.~T., Riot, V.~J., Ritz, S., Robinson,
  S., Roby, W., Roodman, A., Rosing, W., Roucelle, C., Rumore, M.~R., Russo,
  S., Saha, A., Sassolas, B., Schalk, T.~L., Schellart, P., Schindler, R.~H.,
  Schmidt, S., Schneider, D.~P., Schneider, M.~D., Schoening, W., Schumacher,
  G., Schwamb, M.~E., Sebag, J., Selvy, B., Sembroski, G.~H., Seppala, L.~G.,
  Serio, A., Serrano, E., Shaw, R.~A., Shipsey, I., Sick, J., Silvestri, N.,
  Slater, C.~T., Allyn~Smith, J., Chris~Smith, R., Sobhani, S., Soldahl, C.,
  Storrie-Lombardi, L., Stover, E., Strauss, M.~A., Street, R.~A., Stubbs,
  C.~W., Sullivan, I.~S., Sweeney, D., Swinbank, J.~D., Szalay, A., Takacs, P.,
  Tether, S.~A., Thaler, J.~J., Thayer, J.~G., Thomas, S., Thornton, A.~J.,
  Thukral, V., Tice, J., Trilling, D.~E., Turri, M., Van~Berg, R., Vanden~Berk,
  D., Vetter, K., Virieux, F., Vucina, T., Wahl, W., Walkowicz, L., Walsh, B.,
  Walter, C.~W., Wang, D.~L., Wang, S.-Y., Warner, M., Wiecha, O., Willman, B.,
  Winters, S.~E., Wittman, D., Wolff, S.~C., Michael Wood-Vasey, W., Wu, X.,
  Xin, B., Yoachim, P., and Zhan, H.
\newblock {LSST}: From science drivers to reference design and anticipated data
  products.
\newblock \emph{The Astrophysical Journal}, 873\penalty0 (2):\penalty0 111,
  March 2019.
\newblock ISSN 0004-637X.
\newblock \doi{10.3847/1538-4357/ab042c}.
\newblock URL
  \url{https://iopscience.iop.org/article/10.3847/1538-4357/ab042c/meta}.

\bibitem[Kamilov et~al.(2022)Kamilov, Bouman, Buzzard, and
  Wohlberg]{Kamilov2022-ue}
Kamilov, U.~S., Bouman, C.~A., Buzzard, G.~T., and Wohlberg, B.
\newblock {Plug-and-Play} methods for integrating physical and learned models
  in computational imaging.
\newblock \emph{arXiv e-prints}, March 2022.
\newblock URL \url{http://arxiv.org/abs/2203.17061}.

\bibitem[Kidger \& Garcia(2021)Kidger and Garcia]{kidger2021equinox}
Kidger, P. and Garcia, C.
\newblock {E}quinox: neural networks in {JAX} via callable {P}y{T}rees and
  filtered transformations.
\newblock \emph{Differentiable Programming workshop at Neural Information
  Processing Systems 2021}, 2021.

\bibitem[Kingma \& Welling(2013)Kingma and Welling]{Kingma2013-ai}
Kingma, D.~P. and Welling, M.
\newblock {Auto-Encoding} variational bayes.
\newblock \emph{arXiv e-prints}, December 2013.
\newblock URL \url{http://arxiv.org/abs/1312.6114v10}.

\bibitem[Lanusse et~al.(2019)Lanusse, Melchior, and Moolekamp]{Lanusse2019-rx}
Lanusse, F., Melchior, P., and Moolekamp, F.
\newblock Hybrid {Physical-Deep} learning model for astronomical inverse
  problems.
\newblock \emph{arXiv e-prints}, December 2019.
\newblock URL \url{http://arxiv.org/abs/1912.03980}.

\bibitem[Liu et~al.(2007)Liu, Shum, and Freeman]{liu2007face}
Liu, C., Shum, H.-Y., and Freeman, W.~T.
\newblock Face hallucination: Theory and practice.
\newblock \emph{International Journal of Computer Vision}, 75:\penalty0
  115--134, 2007.

\bibitem[Melchior et~al.(2018)Melchior, Moolekamp, Jerdee, Armstrong, Sun,
  Bosch, and Lupton]{melchior2018scarlet}
Melchior, P., Moolekamp, F., Jerdee, M., Armstrong, R., Sun, A.-L., Bosch, J.,
  and Lupton, R.
\newblock Scarlet: Source separation in multi-band images by constrained matrix
  factorization.
\newblock \emph{Astronomy and Computing}, 24:\penalty0 129--142, 2018.

\bibitem[Schawinski et~al.(2017)Schawinski, Zhang, Zhang, Fowler, and
  Santhanam]{Schawinski2017-lv}
Schawinski, K., Zhang, C., Zhang, H., Fowler, L., and Santhanam, G.~K.
\newblock Generative adversarial networks recover features in astrophysical
  images of galaxies beyond the deconvolution limit.
\newblock \emph{Monthly notices of the Royal Astronomical Society},
  467:\penalty0 L110, May 2017.
\newblock ISSN 0035-8711.
\newblock \doi{10.1093/mnrasl/slx008}.
\newblock URL \url{https://ui.adsabs.harvard.edu/abs/2017MNRAS.467L.110S}.

\bibitem[Song et~al.(2020)Song, Sohl-Dickstein, Kingma, Kumar, Ermon, and
  Poole]{song2020score}
Song, Y., Sohl-Dickstein, J., Kingma, D.~P., Kumar, A., Ermon, S., and Poole,
  B.
\newblock Score-based generative modeling through stochastic differential
  equations.
\newblock \emph{arXiv preprint arXiv:2011.13456}, 2020.

\bibitem[Van~Rossum \& Drake~Jr(1995)Van~Rossum and Drake~Jr]{van1995python}
Van~Rossum, G. and Drake~Jr, F.~L.
\newblock \emph{Python reference manual}.
\newblock Centrum voor Wiskunde en Informatica Amsterdam, 1995.

\bibitem[Wang et~al.(2014)Wang, Tao, Gao, Li, and Li]{wang2014comprehensive}
Wang, N., Tao, D., Gao, X., Li, X., and Li, J.
\newblock A comprehensive survey to face hallucination.
\newblock \emph{International journal of computer vision}, 106:\penalty0 9--30,
  2014.

\bibitem[{Yao} et~al.(2020){Yao}, {Gholami}, {Shen}, {Mustafa}, {Keutzer}, and
  {Mahoney}]{yao2021adahessian}
{Yao}, Z., {Gholami}, A., {Shen}, S., {Mustafa}, M., {Keutzer}, K., and
  {Mahoney}, M.~W.
\newblock {ADAHESSIAN: An Adaptive Second Order Optimizer for Machine
  Learning}.
\newblock \emph{arXiv e-prints}, art. arXiv:2006.00719, June 2020.
\newblock \doi{10.48550/arXiv.2006.00719}.

\bibitem[Yu et~al.(2018)Yu, Lin, Yang, Shen, Lu, and Huang]{yu2018generative}
Yu, J., Lin, Z., Yang, J., Shen, X., Lu, X., and Huang, T.~S.
\newblock Generative image inpainting with contextual attention.
\newblock In \emph{Proceedings of the IEEE conference on computer vision and
  pattern recognition}, pp.\  5505--5514, 2018.

\end{thebibliography}
\bibliographystyle{icml2023}

\end{document}